\newcommand{\ket}[1]{\ensuremath{\left|  #1 \right\rangle}}
\newcommand{\bra}[1]{\ensuremath{\left\langle  #1 \right|}}

\newcommand{\ind}[1]{_{\text{#1}}}
\newcommand{\tr}{\text{tr}}

\newcommand\equalhat{\mathrel{\stackon[1.5pt]{=}{\stretchto{%
				\scalerel*[\widthof{=}]{\wedge}{\rule{1ex}{3ex}}}{0.5ex}}}}

\documentclass[aps,prl,reprint,superscriptaddress]{revtex4-2}

\usepackage{amsmath,amssymb,graphicx,color,verbatim,soul}
\usepackage{scalerel,stackengine}
\usepackage{upgreek}
\usepackage{float}
\usepackage{bm}
\usepackage{hyperref}
\hypersetup{colorlinks=true,urlcolor = blue}

\begin{document}

\title{Detecting entanglement structure in continuous many-body quantum systems}
\author{Philipp Kunkel}
\email[]{fieldentanglement@matterwave.de}
\affiliation{Kirchhoff-Institut f\"ur Physik, Universit\"at Heidelberg, Im Neuenheimer Feld 227, 69120 Heidelberg, Germany}
\author{Maximilian Pr\"ufer}
\affiliation{Kirchhoff-Institut f\"ur Physik, Universit\"at Heidelberg, Im Neuenheimer Feld 227, 69120 Heidelberg, Germany}
\author{Stefan Lannig}
\affiliation{Kirchhoff-Institut f\"ur Physik, Universit\"at Heidelberg, Im Neuenheimer Feld 227, 69120 Heidelberg, Germany}
\author{Robin Strohmaier}
\affiliation{Kirchhoff-Institut f\"ur Physik, Universit\"at Heidelberg, Im Neuenheimer Feld 227, 69120 Heidelberg, Germany}
\author{Martin G\"arttner}
\affiliation{Kirchhoff-Institut f\"ur Physik, Universit\"at Heidelberg, Im Neuenheimer Feld 227, 69120 Heidelberg, Germany}
\affiliation{Physikalisches Institut, Universit\"at Heidelberg, Im Neuenheimer Feld 226, 69120 Heidelberg, Germany}
\affiliation{Institut f\"ur Theoretische Physik, Universit\"at Heidelberg, Philosophenweg 16, 69120 Heidelberg, Germany}
\author{Helmut Strobel}
\affiliation{Kirchhoff-Institut f\"ur Physik, Universit\"at Heidelberg, Im Neuenheimer Feld 227, 69120 Heidelberg, Germany}
\author{Markus K.\ Oberthaler}
\affiliation{Kirchhoff-Institut f\"ur Physik, Universit\"at Heidelberg, Im Neuenheimer Feld 227, 69120 Heidelberg, Germany}

\date{\today}

\begin{abstract}

A prerequisite for the comprehensive understanding of many-body quantum systems is a characterization in terms of their entanglement structure.
The experimental detection of entanglement in spatially extended many-body systems describable by quantum fields still presents a major challenge.
We develop a general scheme for certifying entanglement and demonstrate it by revealing entanglement between distinct subsystems of a spinor Bose-Einstein condensate.
Our scheme builds on the spatially resolved simultaneous detection of the quantum field in two conjugate observables which allows the experimental confirmation of quantum correlations between local as well as non-local partitions of the system. 
The detection of squeezing in Bogoliubov modes in a multi-mode setting illustrates its potential to boost the capabilities of quantum simulations to study entanglement in spatially extended many-body systems.

\end{abstract}

\maketitle
\newpage
Entanglement between spatial regions of isolated quantum systems is at the heart of phenomena such as eigenstate thermalization~\cite{DextquotesingleAlessio2016,Deutsch2018} and many-body localization~\cite{Nandkishore2015,Abanin2019}.
Spatial entanglement has been experimentally assessed in small systems with discrete degrees of freedom, such as spins or few particle systems~\cite{Smith2016,Lukin2019,Brydges2019,Bergschneider2019}. These methods rely on the experimental capability of preparing and maintaining pure states as well as detection on the single-particle level.
These requirements are exceedingly hard to fulfill in generic situations in nature where many particles interact and the system is described by continuous quantum fields.
Here we show that the spatially resolved joint measurement of non-commuting field quadratures allows certifying entanglement in these situations with no assumptions about the purity of the global state. 

Our experimental platform is a spin-1 Bose-Einstein condensate (BEC), where spin mixing leads to squeezing in the conjugate spin-1 operators $\hat{S}_x $ and $ \hat{Q}_{yz} $~\cite{Hamley2012}.
To open a more general perspective on entanglement in continuous many-body quantum systems, we define the quantum field $ \hat\Phi = \hat S_x - i\hat Q_{yz} $.
\begin{figure}
	\centering
	\includegraphics{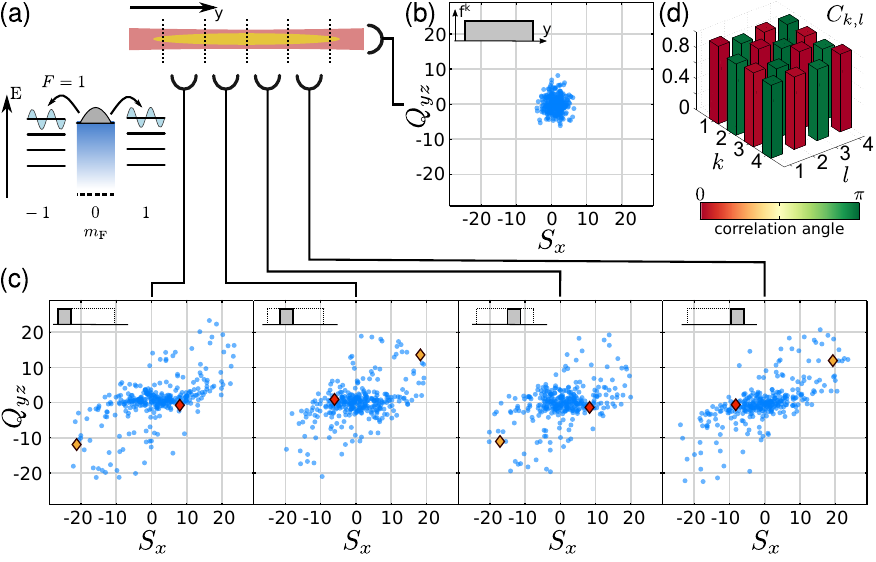}
	\caption{Sampled phase-space distributions of global and local observables: (a)
		Schematics of the level scheme for the physical preparation of the target state and of the longitudinally expanded atomic cloud. The initially populated $ m_\text{F} = 0 $ level is tuned (shading) close to resonance with the third excited eigenmode in $ m_\text{F} = \pm1$ via off-resonant mw dressing. In the right part, the dashed lines indicate the evaluation regions with a length of 20\,$ \upmu $m each.
		(b) and (c) show the sampled phase-space distributions for the global observables and for the local subsystems, respectively.
		The corresponding partitioning functions are shown in the insets.
		The state has been prepared by 800\,ms of spin-mixing dynamics which results in a non-Gaussian distribution in the local partitions.
		The red and yellow points highlight single experimental realizations illustrating the strong correlations between the individual subsystems. These correlations strongly suppress the fluctuations of the global observables leading to a Gaussian distribution as shown in (b).
		(d) shows a quantitative analysis of the first-order coherence between different subsystems revealing strong anti-correlations between neighboring partitions.
	}
	\label{Figure1}
\end{figure}
For our experiments, we initially prepare a BEC of $ ^{87} $Rb atoms in the magnetic substate $ m_\text{F} = 0 $ of the $ F=1 $ hyperfine manifold in a spatially one dimensional situation (see Fig.~\ref{Figure1}(a)). This corresponds to the vacuum state of the quantum field $\hat \Phi(y)$.
By a controlled energy shift of $ m_\text{F} = 0 $ we initiate spin-mixing dynamics. This leads to squeezing of the field quadratures $ (\hat\Phi^\dagger+\hat\Phi)/2 $ and $ (\hat\Phi^\dagger-\hat\Phi)/2i $ and to the build-up of entanglement between spatial subsystems~\cite{Fadel2018,Lange2018,Kunkel2018}.
We directly sample the phase-space distribution of the quantum field via a joint measurement of the two field quadratures, $\hat{S}_x$ and $\hat{Q}_{yz}$. For this, we couple the $ F=1 $ hyperfine manifold to $ F=2 $ which serves as an ancillary system (see \cite{Kunkel2019} and SM for details). After inducing spin rotations via radiofrequency fields, we measure in each experimental realization the populations of the magnetic substates in the $ F=1,2 $ manifolds, from which we estimate the expectation values $ S_x(y) $ and $ Q_{yz}(y) $, respectively.
This estimation can be done with high precision, since approximately $ 600 $ atoms contribute on average to the signal at each position $ y $ after integrating over the transversal directions and the spatial resolution of $ 1.3\,\upmu $m.

In the following, we will show that accessing the local as well as the global phase-space distributions is a key ingredient for entanglement detection.
Entanglement is defined as inseparability of the quantum state with respect to specific partitions of the system.
Given the spatial resolution of our imaging we are able to analyze general partitions via
\begin{align}
	\Phi_k = \mathcal N_k \sum_y f^k(y)\Phi(y)
\label{Eq1}
\end{align}
where the functions $ f^k $ can represent a partitioning in separate spatial regions as well as in mode functions e.g. Bogoliubov modes. Both types will be used in the following.
The normalization $ \mathcal N_k $ is chosen such that the corresponding operators fulfill the commutation relation $ [\hat\Phi_k^{\phantom{\dagger}},\hat\Phi_k^\dagger] = 2 $.
This normalization is possible, since we detect the commutator of the observables $ \hat{S}_x $ and $ \hat{Q}_{yz} $ (for the connection to the detected particle numbers in $ F=1 $ and $ F=2 $ see SM). 

Our readout scheme gives us access to the phase-space distribution in any partition and the direct sampling allows analyzing correlations between them.
To demonstrate this capability, we tune the spin-mixing process into resonance with the third excited mode of the external potential~\cite{Scherer2010,Deuretzbacher2010} and let the system evolve for 800\,ms (see Fig.~\ref{Figure1}(a)). Prior to imaging, we switch off the longitudinal confinement to let the atomic cloud expand by a factor of 4 to an extension of about $ 80\,\upmu $m. The corresponding phase-space distribution of the global observables, i.e. $ f^k(y) = 1 $ is shown in Fig.~\ref{Figure1}(b), which features an isotropic Gaussian distribution.

In Fig.~\ref{Figure1}(c) we show the phase-space distributions for a local analysis corresponding to the partitioning functions as indicated in the insets. These distributions are highly non-Gaussian and characterized by large fluctuations. In conjunction with the observed small fluctuations in the global observables shown in Fig.~\ref{Figure1}(b) this implies strong correlations between the spatial subsystems. 
To reveal the structure of the correlations present we evaluate the first-order coherence
\begin{align}
	C_{k,l} = \frac{\langle \Phi_k^\ast \cdot \Phi_l \rangle}{\sqrt{\langle |\Phi_k|^2\rangle\cdot\langle |\Phi_l|^2\rangle}} \quad k,l \in \{\text{1--4}\} 
\end{align}
where $ \langle\cdot\rangle $ indicates the average over all experimental realizations. $ C  = |C|\mathrm e^{i\theta}$ is in general a complex quantity where the absolute value $ |C| $ quantifies correlations between the local fields and the relative angle between the fluctuations of the local fields is given by $ \theta $.
In Fig.~\ref{Figure1}(d) we find $ \theta $ to be changing by $ \pi $ between neighboring subsystems demonstrating strong anti-correlations as expected from the spatial structure of the populated third excited mode. 
That these correlations are found in any phase-space orientation can be seen explicitly in the examples shown in Fig.~\ref{Figure1}(c), where two realizations in orthogonal phase-space directions are highlighted in red and yellow.

\begin{figure*}
	\centering
	\includegraphics{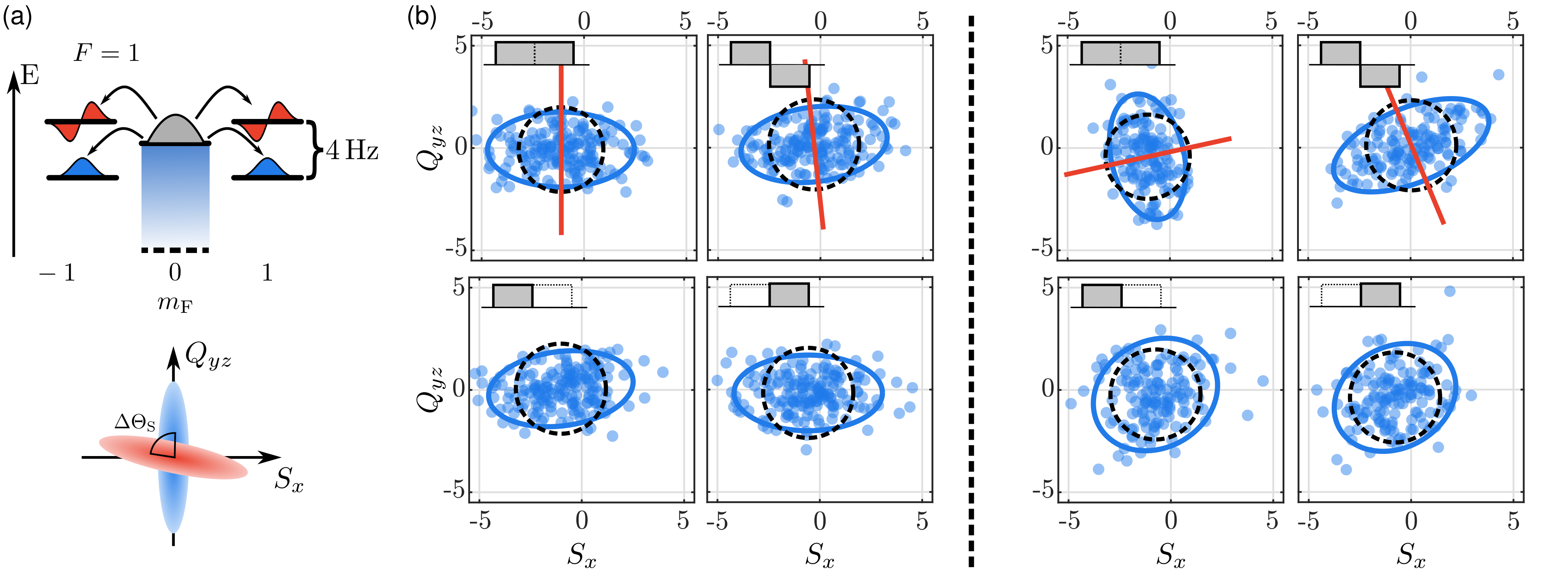}
	\caption{Superposition of two squeezed vacuum states.
		(a) 
		Illustration of simultaneous spin-mixing in two spatial modes (blue and red).
		The initially populated $ m_\text{F} = 0 $ level is tuned (shading) close to resonance with two energetically lowest eigenmodes in $ m_\text{F} = \pm1$ via off-resonant mw dressing (upper part). This results in simultaneous squeezing in the two modes with relative squeezing angle $ \Delta \Theta_\text{S} $ (lower part).
		(b) Phase-space distributions for different partitioning functions as indicated in the insets of each panel. We prepare two different relative squeezing angles of $ \Delta\Theta_\text{S} \approx 0^\circ,90^\circ $ which are shown in the left and right part, respectively. The blue ellipses show the 2 s.d. interval of the distribution and the black dashed circles show the fluctuations expected for the initial vacuum state (including photon shot-noise).
		The upper row reveals the squeezing in the individual modes, where the red lines indicate the axis of smallest fluctuations.
		In the case of $ \Delta\Theta_\text{S} \approx 0^\circ $ we find squeezing in the local partitions while for $ \Delta\Theta_\text{S} \approx 90^\circ $ the fluctuations are increased along all directions compared to the vacuum state (lower row). 
		} 
	\label{Figure2}
\end{figure*}

In the following we will show that knowledge of correlations between quantum fields in different partitions can be used to certify entanglement. For this, we tune the spin-mixing dynamics such that the two energetically lowest trap modes are squeezed as shown in Fig.~\ref{Figure2}(a). We choose an evolution time of 100\,ms in the squeezing regime. Microscopically this corresponds to a mean number of 32 atoms in the magnetic substates $ \pm1 $ compared to the initial BEC of $ 4\times10^4 $ atoms. 
In order to experimentally confirm the expected squeezing of the phase-space distribution in each mode, we choose the partitions indicated in the insets of Fig.~\ref{Figure2}(b).
The corresponding elliptical phase-space distributions are shown in the upper row of Fig.~\ref{Figure2}(b). We compare the standard deviations of the data with the ones expected for the vacuum state as indicated by the blue and black lines, respectively. We find for both partitions fluctuations below the classical limit, where the maximally squeezed orientations are marked by red lines. This confirms the squeezing in the two modes. Due to the energy difference between the two modes, the relative orientation $ \Delta\Theta_\text{S} $ of the two squeezing directions evolves dynamically when switching off the mw dressing of $ m_\text{F} =0 $. 
Employing different hold times we prepare $ \Delta \Theta_\text{S} \approx 0^\circ$ and $90^\circ $.

For a quantitative analysis of the squeezing level the photon shot noise, readout splitting and technical noise that contribute to the measured signal have to be taken into account. 
For the data shown in Fig.~\ref{Figure2} we find after subtracting the photon shot noise contribution minimal fluctuations of $ 0.7\pm0.1 $.
The vacuum fluctuations of the ancillary states limit the level of fluctuations to 0.5 for a perfectly squeezed state. Taking those into account leads to an inferred squeezing of $ -3 $\,dB.
For reaching this detection limit the off-resonant excitations during the splitting pulses have to be suppressed to a level of smaller than $ 10^{-5} $.

Strikingly, even though both spatial modes feature squeezing for the two situations with $ \Delta \Theta_\text{S} \approx 0^\circ$ and $90^\circ $ the spatial entanglement structure differs.
For a relative orientation of $ 0^\circ $ the local phase-space distributions shown in the lower row of Fig.~\ref{Figure2}(b) are similar to the ones found within the mode partitions and also feature fluctuations below the classical limit. In contrast, for an angle of $ 90^\circ $ the local distributions exhibit enhanced fluctuations exceeding those of the vacuum state.
Since this analysis corresponds to a partial trace over the complement of the respective subsystem, finding increased fluctuations in the remaining system is an indication of spatial entanglement.

In order to rigorously show the presence of entanglement we derive an entanglement witness $ \mathcal W $~\cite{Guehne2009} for the experimentally realized joint measurements.
Our measurement strategy allows extracting the field component along arbitrary orientations $ \theta $ via $ F_k(\theta) = \text {Re}(\Phi_k^\dagger\mathrm e^{-i\theta})$ with $ k \in \{\text{L},\text{R}\} $ from the sampled local phase-space distributions, where Re() denotes the real part. 
These quantities allow us to evaluate the following criterion fulfilled for all separable states \cite{Duan2000}:
\begin{equation}
\begin{split}
	\mathcal W =& \Delta^2 u(\theta_\text{L},\theta_\text{R}) + \Delta^2 u(\theta_\text{L}^\prime,\theta_\text{R}^\prime) \\
	& -\left(|\sin(\Delta\theta_\text{L})|+|\sin(\Delta\theta_\text{R})|\right) \geq 0 \,,
\end{split}
	\label{EntanglementBound}
\end{equation}
where $u(\theta_\text{L},\theta_\text{R})=[ F_\text{L}(\theta_\text{L})+F_\text{R}(\theta_\text{R})]/\sqrt{2}$ and $\Delta\theta_\text{k} = \theta_\text{k}-\theta_\text{k}^\prime$. 
Here, $ |\sin(\Delta\theta_\text{k})| $ accounts for the bound of the local uncertainty relation and assumes equal atom numbers in the two partitions (for the imbalanced case see SM). Since the local observables are determined in joint measurements the bound has been adapted with respect to its original form~\cite{Duan2000} by exploiting this knowledge about the measurement process \cite{Arthurs1965,Leonhardt1997}.
The fluctuations in $ u(\theta_\text{L},\theta_\text{R}) $ quantify the degree of correlation between the subsystems and $ \mathcal W<0 $ signals the presence of entanglement.

\begin{figure}
	\centering
	\includegraphics{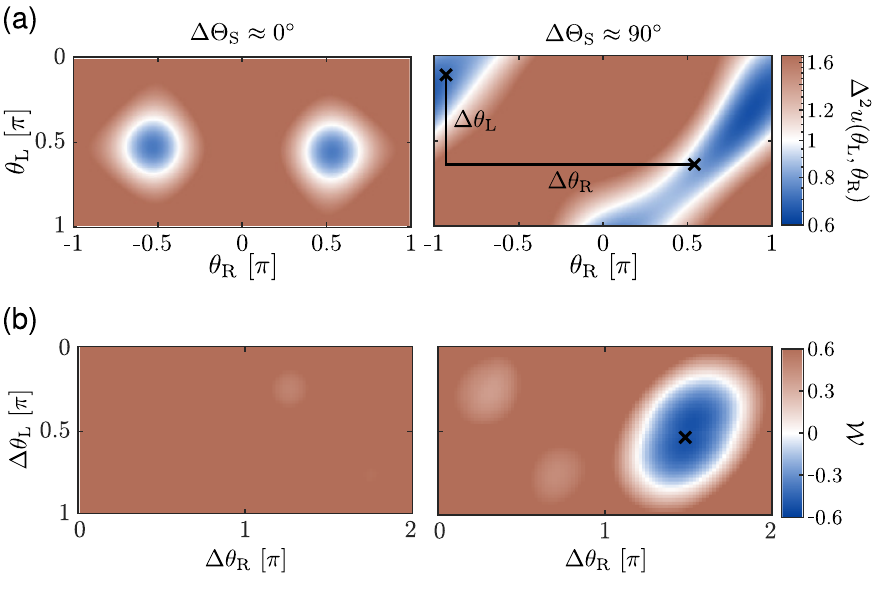}
	\caption{Witnessing entanglement between subsystems.(a) Fluctuations in the observable $ u $ quantifying correlations between the subsystems as a function of the  local field projection angles, $ \theta_\text{L} $ and $ \theta_\text{R} $. Here, the blue regions signal fluctuations below the atomic shot noise limit. While for $ \Delta\Theta_\text{S} = 0^\circ $ (left) we observe fluctuations for specific angle pairs, the continuous diagonal band of reduced fluctuations seen for $ \Delta\Theta_\text{S} = 90^\circ $ (right) is what allows witnessing entanglement.
	(b) Entanglement witness as a function of local relative angles, where the blue regions signal entanglement. For each pair of local relative angles, we minimized the first line of equation \eqref{EntanglementBound}. The point of minimal $ \mathcal W $ is indicated by the black cross. The corresponding pair of local orientations is indicated in (a).}
	\label{Figure3}
\end{figure}

From the sampled phase-space distributions we evaluate the variances $\Delta^2 u(\theta_\text{L},\theta_\text{R})$ for any pair of angles. Figure~\ref{Figure3}(a) shows that for the two relative squeezing angles shown in Fig.~\ref{Figure2} we observe pronounced minima where the fluctuations are suppressed below the atomic shot noise limit ($ \Delta^2 u = 1$). 
We find a violation of Eq.~\eqref{EntanglementBound} by minimizing $\mathcal W$ over all pairs of analysis angles. As the last term in Eq.~\eqref{EntanglementBound} only depends on the relative angles $\Delta\theta_\text{k}$, we minimize the variances with respect to $\theta_\text{k}$ for each choice of $\Delta\theta_\text{k}$. The resulting values of $\mathcal W$ are shown in Fig.~\ref{Figure3}(b), where regions of $\mathcal W<0$ are visible. 
For the case that the squeezing ellipses in the two modes have the same orientation, i.e. $ \Delta \Theta_\text{S} = 0^\circ $, (see Fig.~\ref{Figure2}(b)) we find that the witness does not flag entanglement between the left and right half, although the phase-space distributions of both halves feature squeezing. For $ \Delta \Theta_\text{S} = 90^\circ $ our witness detects entanglement consistent with the enhanced fluctuations in the local partitions shown in Fig.~\ref{Figure2}(b). In this case, we find for the witness a minimal value of  $-0.51\pm0.14$ for $ \Delta\theta_\text{L} = 0.53\,\pi $ and $ \Delta\theta_\text{R} = 1.45\,\pi $, where we subtracted the independently characterized photon shot noise contribution of 0.13 (see SM).
We also find entanglement between the two halves for different relative orientations of the squeezing ellipses, e.g. $ \Delta \Theta_\text{S} = 45^\circ $ (see SM).

\begin{figure}
	\centering
	\includegraphics[width = \columnwidth]{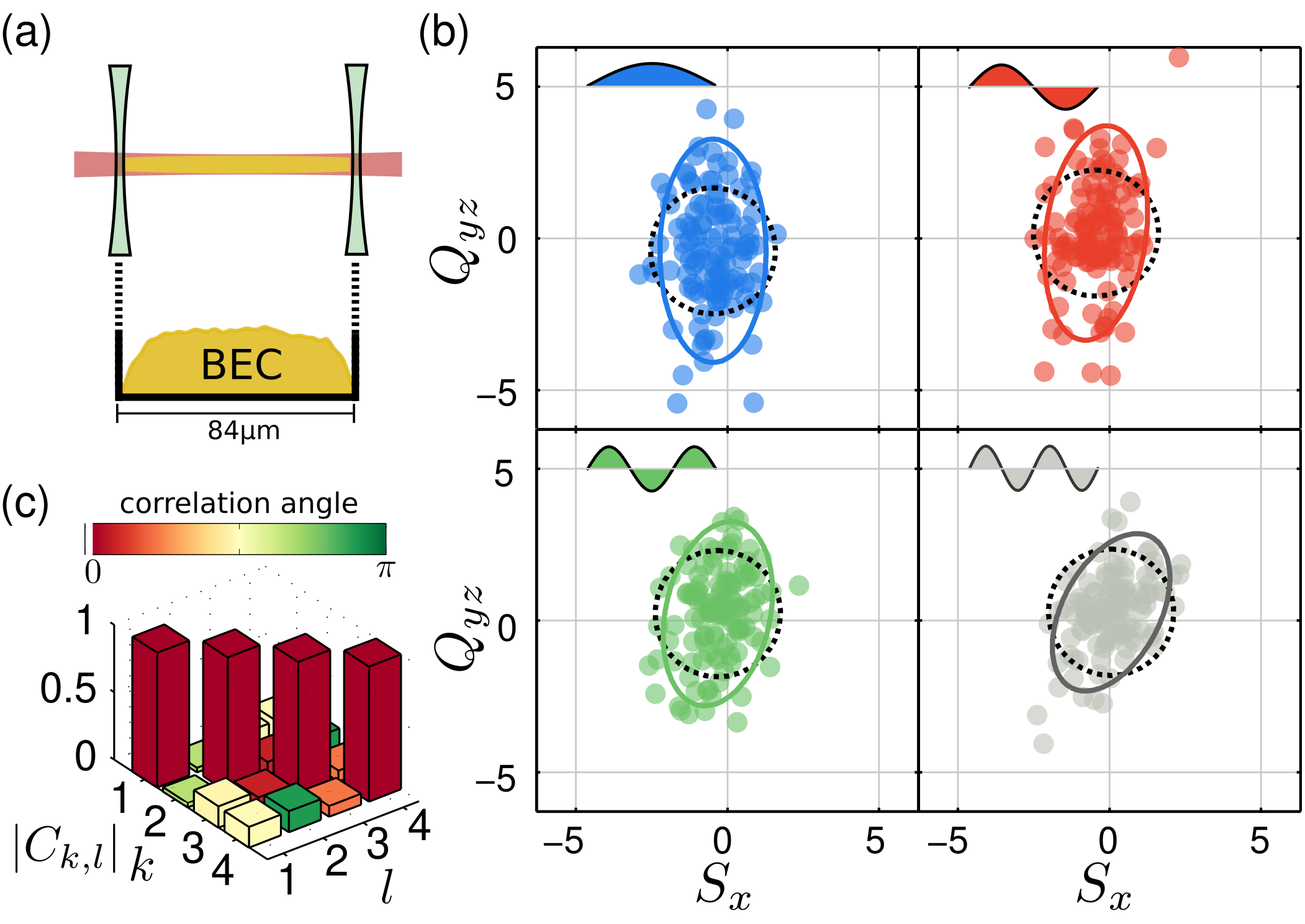}
	\caption{Simultaneous vacuum squeezing of multiple Bogoliubov modes. (a) Schematics of box-like trapping potential implemented by combining an elongated attractive harmonic dipole potential (red) with two repulsive barriers (green). The detected BEC density shown in the lower part (yellow).
		(b) Sampled phase-space distributions for the partitionings indicated in the insets, which reveal vacuum squeezing in the four energetically lowest Bogoliubov modes. The colored solid lines indicate the 2 s.d. interval of the distribution and the dashed black lines show the fluctuations expected for the initial vacuum state including photon shot noise.
		(c) $C_{k,l}$ for the data shown in (b). This confirms the independence of the individual Bogoliubov modes as expected in the low depletion limit.}
	\label{Figure4}
\end{figure}

Having established the experimental capabilities of our method, we now turn to a multimode situation which is the interesting regime for quantum simulation of many-body systems.
Here, we study the quantum structure of multiple spatial modes in a box-like trapping potential (see Fig.\,4).
We confine the atoms to the central part of a weak harmonic trapping potential (with longitudinal and transversal trap frequencies of $ 2\pi\times1.5\, $Hz and $ 2\pi\times170\, $Hz, respectively) by adding two repulsive barriers which are spaced by 84\,$\mu$m. This leads to a flat atomic density as shown in Fig.~\ref{Figure4}(a).
The spin interaction strength of 1.2\,Hz allows us to vacuum squeeze different spatial modes simultaneously since the energy difference between the ground and third excited mode of the box potential is only 1.2\,Hz.

Our experimental observations directly reveal the squeezing in the individual Bogoliubov modes by taking them as partitioning functions $f^k(y)$ (see  Fig~\ref{Figure4}(b) insets).
For a single realization we evaluate the field in mode $k$ according to Eq.~\eqref{Eq1}.
Experimentally, we find squeezing for the four lowest spatial modes by examining the phase space distributions at 100\,ms evolution time (see Fig.~\ref{Figure4}(b)). 
The width of the distributions (shown as coloured ellipses) are smaller than expected for the initially prepared vacuum state (black circle) and amount to a maximal inferred squeezing of $ -8\,$dB .

From the fluctuations along the anti-squeezed axis we infer a mean number of less than 6 atoms in each Bogoliubov mode. Compared to the overall atom number of $\sim 3\times10^4$ this correspond to a regime of low depletion where the Bogoliubov approximation is valid. 
In this regime uncorrelated squeezing dynamics of each mode is expected~\cite{Sau2010}. 
We explicitly confirm that these modes are independent by evaluating the first-order coherence $ C_{k,l} $ (see Eq.~(2)) between the individual partitions as shown in Fig.~\ref{Figure4}(c).
This highlights a major advantage of our detection scheme as extracting these coherences would be experimentally challenging and resource intensive with sequential projective measurements. There mode-selective spin rotations as well as a large number of relative analysis angles would be required.

We present a very general strategy to extract correlations between various partitions of the system as well as their entanglement structure. This lays the ground for resolving the role of entanglement in different phenomena such as thermalization of isolated quantum systems~\cite{Rigol2008}, the emergence of hydrodynamics and quantum effects in gravity with analog quantum simulators~\cite{Boiron2015}. Since our entanglement detection scheme is model independent it may be employed 
for certifying quantum operation of an analog simulator, a task that is indispensable when exploring phenomena beyond the reach of classical devices~\cite{Eisert2020}.

\textbf{Acknowledgments:}
We thank Thorsten Zache for valuable comments on the manuscript and Monika Schleier-Smith, Luca Pezz\'e, Stefan Flörchinger and Jürgen Berges for discussion.
This work is supported by ERC Advanced Grant Horizon 2020 EntangleGen (Project-ID 694561), the Deutsche Forschungsgemeinschaft (DFG, German Research Foundation) under Germany’s Excellence Strategy EXC2181/1-390900948 (the Heidelberg STRUCTURES Excellence Cluster), within the Collaborative Research Center SFB1225 (ISOQUANT) and by the Baden-Württemberg Stiftung gGmbH. {P.K. acknowledges support by the Studienstiftung des deutschen Volkes.}

\setcounter{equation}{0}
\renewcommand{\theequation}{S\arabic{equation}}
\setcounter{figure}{0}
\renewcommand{\thefigure}{S\arabic{figure}}

\clearpage
\onecolumngrid
\section{Supplementary Material}
\subsection{State Preparation}
After condensation the atoms populate the state $ (F,m_\text{F}) = (1,-1) $ in a magnetic field of 1.44\,G, where $ F $ and $ m_\text{F} $ denotes the hyperfine manifold and the magentic substate, respectively. To transfer the atoms into the state $(1,0)$ we employ two resonant mw $ \pi $-pulses coupling the states $ (1,-1)\leftrightarrow(2,0) $ and $ (2,0)\leftrightarrow(1,0) $. To remove residual atoms in $ (1,\pm1) $ we apply a magnetic field gradient, after which we let the magnetic field stabilize again for 100\,ms. 
To remove any atoms in $ F=2 $ and atoms in $ (1,\pm1) $ that might have been generated during the 100\,ms hold time for example via off-resonant spin-mixing, we apply two light pulses which couple the $ F=2 $ manifold to the excited state $ 5^2 $P$ _{3/2} $ and push the atoms via optical forces out of the trap. In between we employ two mw $ \pi $-pulses to transfer atoms from $ (1,\pm1) $ to $ (2,\pm1) $.
 
At a magnetic field of 1.44\,G the second-order Zeeman-shift leads to an energy shift of $ \approx - $150\,Hz of the state $ (1,0) $ with respect to $ (1,\pm1) $. Compared to the spin-spin interaction strength of $ \approx 4\, $Hz, this shift therefore inhibits the creation of atom-pairs in $ (1,\pm1) $.
To initiate spin-mixing we off-resonantly couple the states $ (1,0) $ and $ (2,0) $ with a blue-detuned mw.
This shifts the energy of the state $ (1,0) $ with respect to $ (1,\pm1) $ and allows us to control the spin-mixing process.
\subsection{Trapping potentials}
For the experiments described in the main text, we employ two different trapping geometries.
For the measurements presented in the first three figures, the atoms are confined in a crossed dipole trap consisting of two red-detuned focused laser beams at a wavelength of 1030\,nm. This generates a harmonic confinement with trapping frequencies $ \omega_\parallel = 2\pi\times50 \,$Hz and $ \omega_\perp = 2\pi\times170 \,$Hz in longitudinal and transversal direction, respectively.
In our case, the spatial distribution of the BEC in longitudinal direction is well described by the Thomas-Fermi approximation, which means that the atomic distribution mimics the shape of the external trapping potential up to the chemical potential, which in our experiments is typically about 2\,kHz.
The interactions between the atoms in the states $ (1,0) $ and $ (1,\pm1) $ leads to an effective external potential for the spin-mixing process~\cite{Scherer2010}. For small number of atom pairs in $ (1,\pm1) $ this effective potential is close to harmonic for the two energetically lowest modes and box-like for the energetically higher modes. Compared to the external potential the energy difference in this effective potential is much lower and amounts in our case to about 4\,Hz. Since the spin-mixing dynamics has an interaction strength of about 2--3\,Hz in this geometry, this means that we can tune the pair-creation process selectively into resonance with a single mode or a combination of two external modes.

For the measurements in the multimode regime (Fig.~\ref{Figure4} in the main text), we employ a box-like trapping geometry. For this the atoms are condensed in the cross-dipole trap as before. Subsequently we ramp down one of the confining dipole traps to let the atomic cloud expand in the remaining weakly focused laser beam. Simultaneously, using two blue-detuned laser beams we generate a repulsive potential with a width of about $ 84\,\upmu $m around the center of the harmonic trap.  Over this distance the harmonic potential is approximately constant and, thus, this setup yields a box-like confinement for the atomic cloud. In this geometry the energy levels are given by $\approx 0.08\,\text{Hz}\cdot n^2 $ leading to an energy difference between the ground and third excited mode of $ \approx 1.2\, $Hz. 
This is also the relevant energy spacing for spin-mixing as in the Thomas-Fermi approximation the effective potential is again a box with a width of 84$ \,\upmu $m. In this trap geometry the interaction strength is typically 1.2\,Hz, which means that multiple spatial modes can be tuned close to resonance simultaneously.

\subsection{Joint measurement of two conjugate field components}
\begin{figure}
	\centering
	\includegraphics{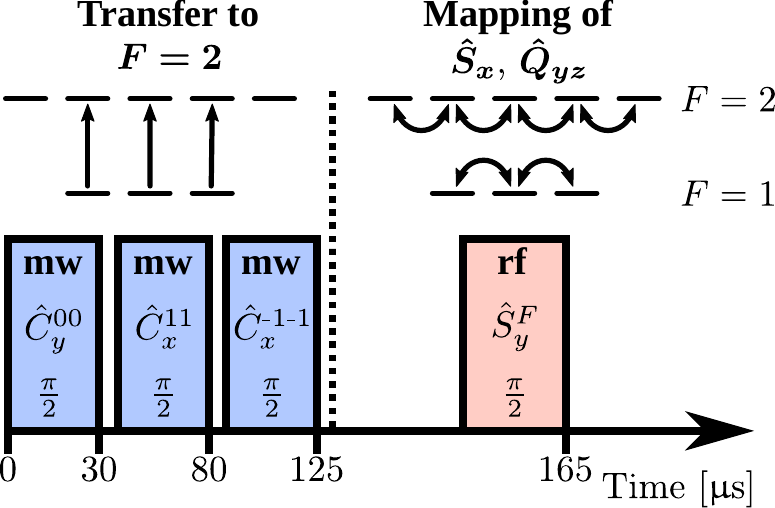}
	\caption{Sketch of the readout sequence for the joint measurement of $ S_x $ and $Q_{yz}$}
	\label{Supp:ReadSeq}
\end{figure}

For the definition of the conjugate spin-1 operators $ \hat{S}_x $ and $ \hat{Q}_{yz} $ we choose the one given in~\cite{Hamley2012}. Note that in the main text we use a different normalization. These two observables are given in second quantization as
\begin{equation}
\begin{aligned}
	\hat{S}_x(y) &= \frac{1}{\sqrt{2}}\,\hat{a}^\dagger_{1,0}(y)\left[\hat{a}_{1,+1}(y)+\hat{a}_{1,-1}(y)\right]\; +\; \text{h.c.} \\
	\hat{Q}_{yz}(y) &= \frac{i}{\sqrt{2}}\,\hat{a}^\dagger_{1,0}(y)\left[\hat{a}_{1,+1}(y)+\hat{a}_{1,-1}(y)\right]\; +\; \text{h.c.}
\end{aligned}
\end{equation}
where $ \hat{a}^{(\dagger)}_{(F,m_\text{F})}(y) $ is the annihilation (creation) operator in the state $ (F,m_\text{F}) $ and at position $ y $ and h.c. denotes the Hermitian conjugate. These two operators are often referred to as spin and quadrupole components and they fulfill the commutation relation
\begin{equation}
	[\hat{S}_x,\hat{Q}_{yz}] = -i [2\hat{N}_0 - (\hat{a}_{(1,+1)}^\dagger+ \hat{a}_{(1,-1)}^\dagger)(\hat{a}_{(1,+1)} + \hat{a}_{(1,-1)})]
\end{equation}
where the operator $ \hat{N}_0 $ denotes the particle number operator in the state $ (1,0) $. In the limit of small depletion, the commutator is approximated by $-i\,2n_\text{tot} $, where $ n_\text{tot} $ is the total atom number.

For a joint measurement of these two conjugate observables we employ the following sequence, illustrated in Fig.~\ref{Supp:ReadSeq}. After expansion of the atomic cloud we use three mw $ \pi/2 $-pulses to couple the magnetic substates $ m_\text{F} = 0,\pm1 $ in $ F=1 $ to the corresponding substates in the $ F=2 $ manifold. With this, we transfer on average half of the atoms to the $ F=2 $ hyperfine manifold, which serves as an auxiliary system for the readout and with that allows the simultaneous extraction of two conjugate observables in a single experimental realization.

As we are interested in the transversal spin and quadrupole degree of freedom we use a radio-frequency (rf) pulse, which, on a spin sphere, implements a $ \pi/2 $-rotation. The frequency of this rf pulse is chosen such that it has the same detuning in the $ F=1 $ and the $ F=2 $ hyperfine manifold.
At the chosen magnetic field of 1.44\,G, the resonance frequencies in the two hyperfine manifolds differ by 4.4\,kHz, leading to a detuning of the rf pulse by 2.2\,kHz. This is more than five times smaller than the employed resonant Rabi frequency of 12\,kHz, which means that the rf-pulse can be used to implement a spin rotation corresponding to a resonant $ \pi/2 $-pulse in each manifold.

In $ F=1 $ this rotation maps the observable $ \hat{S}_x $ onto the measurable populations in the states $ (1,\pm1) $. 
The relative phases of the mw pulses determine which observable is mapped onto the populations in the $ F=2 $ manifold. By changing the phase of the mw pulse coupling the states $ (1,0) \leftrightarrow (2,0) $ by $ \pi/2 $ we can change the observable from the spin component $ \hat S_x $ to the quadrupole component $ \hat Q_{yz} $. We independently calibrated the phases of the mw pulses such that after the readout sequence we extract in each partition the two observables from the measured populations via
\begin{equation}
\begin{aligned}
	S_x^k = \frac{\sum_y f^k(y) [n_{(1,+1)}(y)-n_{(1,-1)}(y)]}{\sqrt{\sum_y f^k(y)^2 [n_{(1,+1)}(y)+n_{(1,-1)}(y)]}} \equalhat \frac{\hat{S}_x^k}{\sqrt{2n_\text{tot}^k}}\\
	Q_{yz}^k = \frac{\sum_yf^k(y)[ n_{(2,+2)}(y)-n_{(2,-2)}(y)]}{\sqrt{\sum_y f^k(y)^2 [n_{(2,+2)}(y)+n_{(2,-2)}(y)]}} \equalhat \frac{\hat{Q}_{yz}^k}{\sqrt{2n_\text{tot}^k}}
\end{aligned}
\label{Eq:SuppExtraction}
\end{equation}
where $ n_{(F,m_\text{F})}(y) $ is the measured atom number in the state $ (F,m_\text{F}) $ at position $ y $ and $ n_\text{tot}^k $ denotes the total atom number in partition $ k $. 
We define the quantum field $\hat \Phi$ via these observables, which are normalized to the atom number found in each partition, via $ \Phi_k = S_x^k-iQ_{yz}^k$, such that the corresponding operator obeys the commutation relation:
\begin{equation}
\begin{aligned}
	\left[\hat{\Phi}_k^{\phantom{\dagger}},\hat{\Phi}_k^\dagger\right] &= \left[\frac{\hat{S}_x^k}{\sqrt{2n_\text{tot}^k}}-i\frac{\hat{Q}_{yz}^k}{\sqrt{2n_\text{tot}^k}},\frac{\hat{S}_x^k}{\sqrt{2n_\text{tot}^k}}+i\frac{\hat{Q}_{yz}^k}{\sqrt{2n_\text{tot}^k}}\right]\\
	& = \frac{i}{n_\text{tot}}\left[\hat{S}_x^k,\hat{Q}_{yz}^k\right]\\
	& = 2.
\end{aligned}
\end{equation}

\subsection{Measured squeezing values}

In Table 1, we provide the measured fluctuations and inferred squeezing values for the measurements shown in the main text. 
In order to infer the squeezing, we take into account the effect of the joint measurement, which is described in detail in~\cite{Kunkel2019}. In our case, the measured minimal fluctuations are connected to the inferred squeezing value $ \xi_k $ via
\begin{equation}
	\xi_k = 10\cdot \log_{10}\left[2\left(\Delta^2 F_k(\theta_\text{min})-0.5\right)\right]	
\end{equation}

\begin{table}[]
	\begin{tabular}{|c|c|c|c|}
		\hline
		Configuration & Partition & Measured Minimal Fluctuations & Inferred Squeezing \\ \hline
		harmonic trap, 	&  symmetric (left+right)  &  $ 0.73\pm 0.09 $ & $ -3.4\substack{+1.4\\-2.2}\, $dB \\ 
		$ \Delta\Theta_\text{S} = 90^\circ $	& antisymmetric (left-right) & $ 0.77\pm0.12 $ &  $ -2.7\substack{+1.6\\-2.6} \,$dB \\ 
			& left half &   $ 1.63\pm0.17 $  & $ +3.5\substack{+0.6\\-0.7} \,$dB \\ 
			& right half & $ 1.45\pm0.17 $ & $ +2.8\substack{+0.7\\-0.9}\,$dB\\ \hline
		harmonic trap,  & symmetric (left+right) & $ 0.75\pm0.08 $ & $ -3\substack{+1.2\\-1.7} \,$dB \\
		$ \Delta\Theta_\text{S} = 0^\circ$ & antisymmetric (left-right) & $ 0.74\pm0.09 $ & $-3.2\substack{+1.4\\-2.0} \,$dB\\
		& left half & $ 0.73\pm0.08 $ & $ -3.4\substack{+1.3\\-1.9} \,$dB \\
		& right half & $ 0.77\pm0.09 $ & $ -2.7\substack{+1.2\\-1.8} \,$dB \\ \hline
		box trap & ground mode &$  0.71\pm0.10$ & $-3.8\substack{+1.7\\-2.8} \,$dB\\
		& first excited mode &$ 0.62\pm0.09 $ & $-6.2\substack{+2.4\\-6.0} \,$dB \\
		& second excited mode &$ 0.68\pm0.10 $ & $-4.4\substack{+1.9\\-3.5} \,$dB\\
		& third excited mode & $ 0.58\pm0.07 $&  $-8.0\substack{+2.7\\-9.0} \,$dB\\ \hline
		
	\end{tabular}
\caption{Measured fluctuations and inferred squeezing}
\end{table}
\subsection{Entanglement verification for $ \Theta_\text{S} = 45^\circ $}
\begin{figure}
	\centering
	\includegraphics{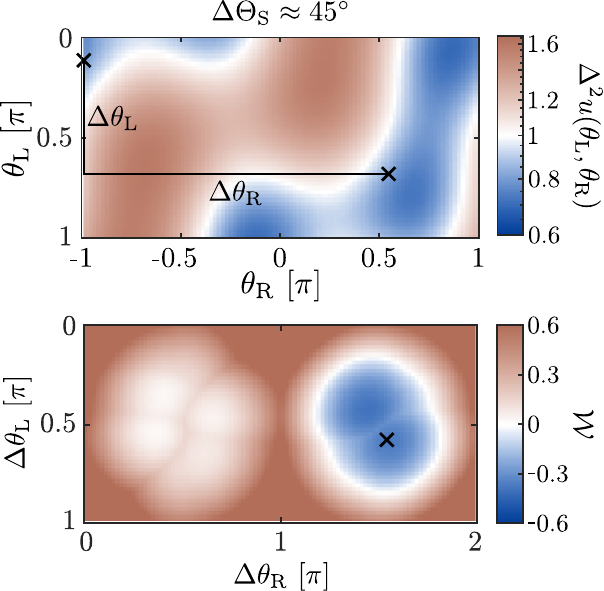}
	\caption{Entanglement witness. Same as Fig.~\ref{Figure3} in the main text with 50\,ms of evolution time.}
	\label{Supp:Entanglement}
\end{figure}
In addition to the two relative squeezing angles described in the main text, we prepare a relative squeezing angle $ \Theta_\text{S} = 45^\circ $. Here, this is achieved via a shorter evolution time of the spin-mixing dynamics of 50\,ms. The corresponding plots showing the correlations and the witness for entanglement between the left and right half of the atomic cloud are shown in Fig.~\ref{Supp:Entanglement}.
In this situation, we measure for the  entanglement witness given in Eq.~\eqref{EntanglementBound} $ \mathcal W = -0.39 \pm0.14 $ certifying entanglement between the left and right half.

\subsection{Imaging Calibration}
\begin{figure*}
	\centering
	\includegraphics{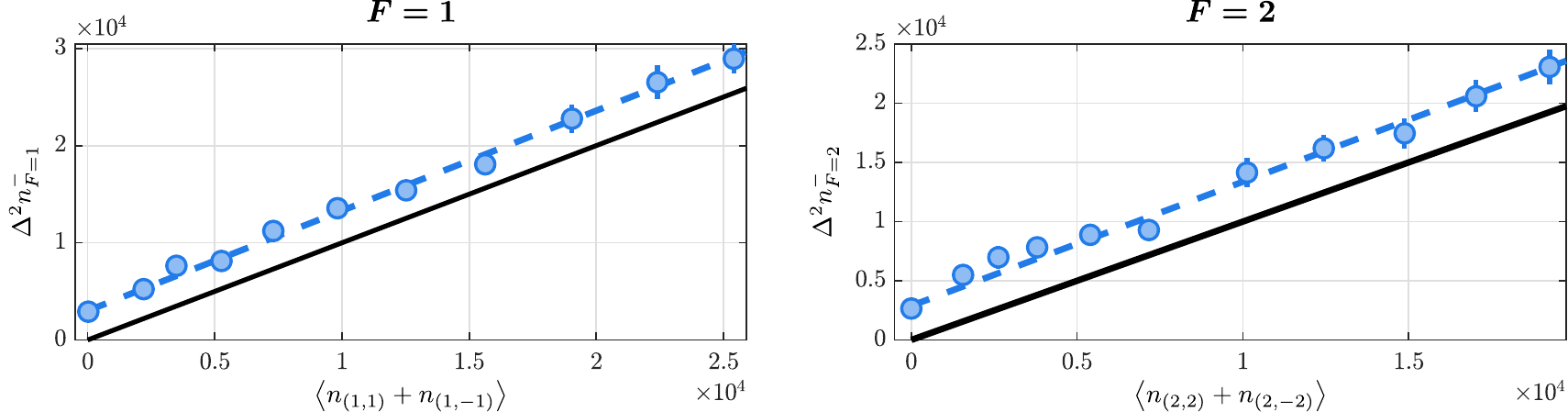}
	\caption{Imaging Calibration. The left and right panels show the atom number fluctuations as a function of the total atom number measured in $ F=1 $ and $ F=2$, respectively. The dashed lines are linear fits to the data while the black lines are the theoretical expectation for a coherent state without technical fluctuations.}
	\label{Supp:ImagCal}
\end{figure*}

Details about our imaging system and the calibration procedure are reported in \cite{Muessel2013}.
To reduce imaging noise we employ a fringe removal algorithm as detailed in~\cite{Ockeloen2010}.

To check the calibration of our imaging we prepare a coherent spin state with approximately equal mean atom numbers in the states $ (1,\pm1) $ and $ (2,\pm2) $, respectively.
Starting from the state $ (1,-1) $ we use two mw $ \pi $-pulses coupling the states $ (1,-1)\leftrightarrow(2,0) $ and $ (2,0)\leftrightarrow(1,0) $.
Analogous to the experimental sequence described above we carefully clean any residual atoms in the states $ (1,\pm 1)$ and in the $ F=2 $ manifold. The only difference here is that we apply the mw together with the cleaning light pulses after the expansion of the atomic cloud. This ensures that the last cleaning is performed as close as possible to the readout pulses and at low atomic density which minimizes the probability for transferring atoms to $ (1,\pm1) $ via collisional interactions.
After this cleaning, we use a mw $ \pi/2 $-pulse to generate an equal superposition of the states $ (1,0) $ and $ (2,0) $.
Subsequently, an rf $ \pi/2 $-pulse is used to prepare an equal superposition of the states $ (1,\pm1) $ and $ (2,\pm2) $. In $ F=2 $ a small fraction, i.e. 12.5\%, of the initial atom number remains in the state $ (2,0) $ after the rf pulse.
To control the total atom number, we vary the height of the external potential during condensation.

In order to mitigate technical noise contributions, we divide the atomic signal into two halves and extract the atom number difference $ n_{F}^{-,\text{L/R}} = n^\text{L/R}_{(F,+F)}-n^\text{L/R}_{(F,-F)}$ in each half and for each manifold $ F=1,2 $.
We subtract the value of the right half from the one of the left to obtain $ n_F^{-} = n_{F}^{-,\text{L}} -n_{F}^{-,\text{R}} $.
For each setting of the atom number we compute the variance $ \Delta^2 n_F^{-} $ and plot it as a function of the measured mean atom number $ \langle n_{(F,+F)}+n_{(F,-F)}\rangle $ in the respective manifold as shown in Fig.~\ref{Supp:ImagCal}.
For a coherent state one expects to find multinomial fluctuations of the populations implying $ \Delta^2 n_{F=1,2}^{-} =  \langle n_{(F,+F)}+n_{(F,-F)}\rangle $.

From a fit to the data we extract a slope of $ 1.03\pm0.03 $ for $ F=1 $ and a slope of $ 1.05 \pm 0.04 $ for $ F=2 $ which is consistent with coherent state fluctuations.
For the offset we find $ 3,000\pm170 $ for $ F=1 $ and $ 2,900 \pm 240 $ for $ F=2 $. These values include the photon shot noise contribution of $ 1,500 $ for $ F=1 $ and $ 1,300 $ for $ F=2 $ which we compute via Gaussian error propagation from the number of detected photons.

\subsection{Entanglement Witness}

In this supplementary section we give a proof of Eq.~\eqref{EntanglementBound} of the main text. We drop operator hats here for notational convenience.

\textbf{Inseparability criteria based on variance sums:} We first recall the inseparability criterion reported in Ref.~\cite{Duan2000} using a notation which will be convenient in the following. We will subsequently adapt this criterion to the case of jointly measured observables.
Let
\begin{align}
u &=  O_{1,\text{L}} + O_{1,\text{R}} \\
v &=  O_{2,\text{L}} + O_{2,\text{R}}
\end{align}
where $O_{i,k}$ are Hermitian operators (observables) acting on subsystem $k$ of a bipartite system. Any separable state $\rho=\sum_i p_i \rho_{i,\text{L}}\otimes \rho_{i,\text{R}}$ obeys
\begin{align}
\Delta^2 u + \Delta^2 v & = \sum_i p_i \left(\langle u^2\rangle_i + \langle v^2\rangle_i  - \langle u\rangle_i^2 - \langle v\rangle_i^2 \right) \label{eq:duan1} \\
& = \sum_i p_i \left(\langle O_{1,\text{L}}^2\rangle_i + 2 \langle O_{1,\text{L}}\rangle_i \langle  O_{1,\text{R}}\rangle_i +  \langle  O_{1,\text{R}}^2\rangle_i +\langle O_{2,\text{L}}^2\rangle_i + 2 \langle O_{2,\text{L}}\rangle_i \langle  O_{2,\text{R}}\rangle_i + \langle  O_{2,\text{R}}^2\rangle_i  
- \langle u\rangle_i^2 - \langle v\rangle_i^2\right) \label{eq:duan2} \\
& =\sum_i p_i \left(\Delta^2 O_{1,\text{L}} + \Delta^2 O_{1,\text{R}} + \Delta^2 O_{2,\text{L}} +\Delta^2 O_{2,\text{R}} \right) \label{eq:duan3}  \\
& \qquad + \sum_i p_i \langle u \rangle_i^2 - \left( \sum_i p_i \langle u \rangle_i \right)^2 + \sum_i p_i \langle v \rangle_i^2 -  \left(\sum_i p_i \langle v \rangle_i \right)^2 \label{eq:duan4}  \\
& \geq \sum_i p_i \left(\Delta^2 O_{1,\text{L}} + \Delta^2 O_{2,\text{L}} + \Delta^2 O_{1,\text{R}} +\Delta^2 O_{2,\text{R}} \right) \label{eq:duan5}  \\
&\geq \sum_i p_i \left( |\langle [O_{1,\text{L}}, O_{2,\text{L}}] \rangle_i| + |\langle [O_{1,\text{R}}, O_{2,\text{R}}] \rangle_i| \right) \label{eq:duan6}  \\
&= |\langle [O_{1,\text{L}}, O_{2,\text{L}}] \rangle_\rho| + |\langle [O_{1,\text{R}}, O_{2,\text{R}}] \rangle_\rho| \,. \label{eq:duan7} 
\end{align}
In this derivation we used the definition of the separable state $\rho$ and of the variance in line \eqref{eq:duan1}. The index $i$ in the expectation value means that it is taken with respect to the product state $\rho_i=\rho_{i,\text{L}}\otimes \rho_{i,\text{R}}$, and the index $\rho$ denotes an expectation value with respect to the full state. In line \eqref{eq:duan2} we used that $\rho_i$ is a product state (i.e. we used the separability of $\rho$) to factorize $\langle O_{1,\text{L}}  O_{1,\text{R}}\rangle_i = \langle O_{1,\text{L}}\rangle_i \langle  O_{1,\text{R}}\rangle_i$. Lines \eqref{eq:duan3} and \eqref{eq:duan4} are simple rearrangements using $2\langle O_{1,\text{L}}\rangle_i \langle  O_{1,\text{R}}\rangle_i = \langle u\rangle_i^2 - \langle O_{1,\text{L}}\rangle_i^2 - \langle  O_{1,\text{R}}\rangle_i^2$. To obtain \eqref{eq:duan5} we used that line \eqref{eq:duan4} is bounded below by zero due to the Cauchy Schwarz inequality $\mathbf{x}^2\mathbf{y}^2 \geq |\mathbf{x}\cdot \mathbf{y}|^2$ for the vectors $x_i = \sqrt{p_i}$ and $y_i=\sqrt{p_i}|\langle u\rangle_i|$, which gives $\sum_i p_i \langle u \rangle_i^2 = \left( \sum_i p_i\right) \left(\sum_i p_i \langle u \rangle_i^2 \right) \geq \left| \sum_i p_i | \langle u \rangle_i | \right|^2 \geq \left( \sum_i p_i \langle u \rangle_i \right)^2 $. Next, in line \eqref{eq:duan6}, we used that from Heisenberg's uncertainty relation it follows that $\Delta^2 O_{1,\text{L}} + \Delta^2 O_{2,\text{L}} \geq 2 \Delta O_{1,\text{L}} \Delta O_{2,\text{L}} \geq |\langle [O_{1,\text{L}}, O_{2,\text{L}}] \rangle_i| $. At this point the derivation below will use that for jointly measured observables a modified uncertainty relation applies. In the last step $\eqref{eq:duan7}$ we used that the expectation values of the commutators are state independent. In the original publication \cite{Duan2000} the authors considered the quadrature operators $x$ and $p$, for which this is satisfied. In our case it is not a priori the case. However, when working in the regime of undepleted state $ m_\text{F} = 0 $, the quantities $S_x$ and $Q_{yz}$ map to $x$ and $p$ in the sense that their commutator becomes state independent, which justifies this step.

\textbf{Generalized Arthurs Kelly uncertainty relation:}
We now show that for observables measured jointly by coupling the system to a set of ancillary modes, the right hand side of the uncertainty relation increases by a factor of two.

The inseparability criterion discussed above assumes that the observables $O_{1}$ and $O_{2}$ are measured in independent runs of the experiment. Two incompatible observables, i.e.\ observables that do not commute or are not diagonal in the same basis, like position and momentum, cannot be measured simultaneously in the sense that the probability distribution over the measurement outcomes of both observables cannot be determined in a single measurement setting. When considering generalized measurements, physically realized by the coupling to meter systems (or auxiliary systems), it turns out that the means (expectation values) of incompatible observables can be determined simultaneously but the probability distribution of the outcomes becomes broadened compared to the original operators, i.e. what we measure is always an observable that approximates the original one with limited precision \cite{Muynck1983}. In other words a \emph{joint} measurement always approximates incompatible observables as it introduces errors with respect to the ideal measurement of each of them \cite{Branciard2013}. This limited precision is what leads to a modified uncertainty relation for jointly measured observables. We will now formally derive an uncertainty relation for a specific setting with bosonic modes following the arguments of Arthurs and Kelly \cite{Arthurs1965}. This result will then be used in the bipartite case to derive the entanglement witness for jointly measured observables.
We remark that it has been shown that jointly measured observables cannot be used to exclude local hidden-variable theories~\cite{Quintino2014, Uola2014}. 

We consider a system on $n$ bosonic modes (system modes) which are supplemented by $m$ ancillary modes. We prepare the system modes in a certain state $\rho_s$ and the ancillary modes are initially empty. Thus the total system is in state $\rho = \rho_s \otimes \ket{0}\bra{0}^{\otimes m} $. The readout consists in a sequence of unitary operations $U_R^{\phantom{\dagger}}=\prod_k U_R^{(k)}$, with $U\ind{R}^{(k)}=\exp[-it_k H\ind{R}^{(k)}]$ generated by Hamiltonians of the form $H\ind{R}^{(k)}=\sum_{\alpha\beta}h^{(k)}_{\alpha\beta}a_\alpha^\dagger a_\beta^{\phantom{\dagger}}$ where $\alpha$, $\beta$ label system and ancillary modes, and subsequent detection of the occupations of all the modes.
From the measured probability distribution $P(N_1,\ldots, N_{n+m})$ we can extract observables $\langle \tilde{O}\rangle = \langle \sum_\alpha e_\alpha N_\alpha\rangle $ and their moments.
We consider a pair of observables $\tilde{O}_1$, $\tilde{O}_2$ of this form, which means that their expectation values are first moments of the measured distribution. The distribution of measured outcomes, $P$, provides an approximation to two non-commuting observables $O_1$, $O_2$ ($[O_1,O_2]=O_3\neq 0$), defined below, which act on the system modes only.
With the state after applying the readout operation $\rho_f = U_R \rho U_R^\dagger$ we can write expectation values as
\begin{equation}
\label{eq:exp_val_POVM}
\langle \tilde{O}\rangle_{\rho_f} = \tr (\tilde{O} U_R^{\phantom{\dagger}} \rho U_R^\dagger) = \tr ( U_R^\dagger \tilde{O} U_R^{\phantom{\dagger}} \rho) = \langle O+\delta O \rangle_{\rho} = \langle O \rangle_{\rho} \,.
\end{equation}
Since $H_R^{(k)}$ are quadratic $U_R^{\phantom{\dagger}}$ enacts a linear transformation of the operators $a_\alpha$ and thus $U_R^\dagger \tilde{O} U_R^{\phantom{\dagger}}$ is still quadratic in the mode operators. Here we have defined $O$ as the part containing operators acting on system modes only and $\delta O$ as the part containing terms that are at least linear in auxiliary mode operators.  
Since the auxiliary modes are unoccupied before the readout is applied, the expectation value of $\delta O $ with respect to $\rho$ vanishes, which we used in the last step in Eq.~\eqref{eq:exp_val_POVM} to show that the mean of $\tilde{O}$ equals the mean of $O$. Thus, the approximate measurement using the auxiliary modes yields the correct means but $\delta O $ accounts for the "error" introduced by the joint measurement, which manifests in an increased variance of $U\ind{R}^\dagger\tilde{O} U\ind{R}^{\phantom{\dagger}}$ compared to $O$. In the following we will make repeated use of the fact that expectation values with respect to $\rho$ of operators linear in $\delta O$ vanish.

The commutator of two jointly measured observables $\tilde{O}_1$ and $\tilde{O}_2$ vanishes as jointly measured observables have to be diagonal in the same basis by definition (they are just linear combinations of number operators in our case). Therefore we have
\begin{equation}
0= \langle [\tilde{O}_1,\tilde{O}_2] \rangle_{\rho_f} = \langle [O_1+\delta O_1,O_2+\delta O_2] \rangle_{\rho} = \langle [O_1,O_2] \rangle_{\rho} + \langle [\delta O_1,\delta O_2] \rangle_{\rho}
\end{equation}
and thus $|\langle [O_1,O_2] \rangle_{\rho}| =|\langle [\delta O_1,\delta O_2] \rangle_{\rho}|$.
With this the uncertainty relation for jointly measured observables becomes
\begin{align}
\Delta^2\tilde{O}_1 \Delta^2\tilde{O}_2 &=  \left( \langle \tilde{O}_1^2 \rangle_{\rho_f} - \langle \tilde{O}_1 \rangle_{\rho_f}^2 \right) \left( \langle \tilde{O}_2^2 \rangle_{\rho_f} - \langle \tilde{O}_2 \rangle_{\rho_f}^2 \right) \\
& = \left( \langle (O_1+\delta O_1)^2 \rangle_{\rho} - \langle O_1+\delta O_1 \rangle_{\rho}^2 \right) \left( \langle (O_2+\delta O_2)^2 \rangle_{\rho} - \langle O_2+\delta O_2 \rangle_{\rho}^2 \right)  \\
& =  \left( \Delta^2 O_1 + \Delta^2\delta O_1 \right)\left( \Delta^2 O_2 + \Delta^2\delta O_2 \right) \label{eq:AKineq1}\\
& \geq \left( \Delta O_1 \Delta O_2 +  \Delta \delta O_1 \Delta\delta O_2 \right)^2 \label{eq:AKineq2} \\
& \geq     \left( \frac{1}{2}|\langle [O_1,O_2]\rangle_\rho |  +  \frac{1}{2}|\langle [\delta O_1,\delta O_2]\rangle_\rho | \right)^2 \\
& = |\langle [O_1,O_2]\rangle_\rho |^2
\end{align}
Here we used that the ancillary modes are empty for $\rho$ to obtain line \eqref{eq:AKineq1}. Line \eqref{eq:AKineq2} is an elementary inequality, and in the last line we used the equality of the absolute values of the commutators derived above. The added fluctuations due to the joint measurement manifest as the additional fluctuations due to the ancillary modes in Eq.~\eqref{eq:AKineq2}.

\textbf{Inseparability criterion and application to the spinor BEC case:}
Applying the resulting uncertainty relation, $\Delta^2 \tilde{O}_{1,k}\Delta^2 \tilde{O}_{2,k} \geq|\langle [O_{1,k},O_{2,k}]\rangle_\rho |^2$, in the derivation of the inseparability criterion of Duan et al.~\cite{Duan2000} we obtain the following criterion for jointly measured observables: If $\rho$ is separable, then
\begin{equation}
\label{eq:insep_AK}
\Delta^2 u + \Delta^2 v \geq 2\left( |\langle [O_{1,\text{L}},O_{2,\text{L}}]\rangle_\rho| + |\langle [O_{1,\text{R}},O_{2,\text{R}}]\rangle_\rho| \right) \,.
\end{equation}

To certify entanglement between the right and left half as described in the main text, we choose $ u =[ F_\text{L}(\theta^{\phantom{\prime}}_\text{L})+  F_\text{R}(\theta^{\phantom{\prime}}_\text{R})]/\sqrt{2}$ and $ v =[ F_\text{L}(\theta^{\prime}_\text{L})+  F_\text{R}(\theta^{\prime}_\text{R})]/\sqrt{2}$. In terms of the extracted quantities in $ F=1,2 $, these field components along $ \theta $ are explicitly given in Eq.~\eqref{Eq:SuppExtraction} via $ F_k(\theta)=\text{Re}(\Phi_k\mathrm e^{-i\theta})= \cos(\theta) S_x^k + \sin(\theta)Q_{yz}^k $. This means that the relevant operators in eq.~\eqref{eq:insep_AK} are
\begin{equation}
  O_{1(2),k} = \frac{1}{2\sqrt{n^k_\text{tot}}}\left(\cos(\theta_k^{(\prime)}) \hat{S}_x^k + \sin(\theta_k^{(\prime)})\hat{Q}_{yz}^k\right).
\end{equation}
Thus, the relevant commutation relations are
\begin{equation}
\begin{aligned}
	{[O_{1,k},O_{2,k}]} &= \frac{1}{4n^k_\text{tot}}\left[\cos(\theta_k) \hat{S}_x^k + \sin(\theta_k)\hat{Q}_{yz}^k,\cos(\theta_k^{\prime}) \hat{S}_x^k + \sin(\theta_k^{\prime})\hat{Q}_{yz}^k\right]\\
	 &= -\frac{i}{2} \sin(\theta_k-\theta_k^\prime).
\end{aligned}
\end{equation}
In the last step we again used $ n_\text{tot}^k \approx n_0^k $, which is well fulfilled for weakly squeezed states.
Inserting this into Eq.~\eqref{eq:insep_AK} yields Eq.~\eqref{EntanglementBound} in the main text.

\end{document}